\documentstyle[floats,aps,preprint]{revtex}
\begin{document}
\draft
\preprint{\vbox{ \hbox{SOGANG-HEP 261/99} \hbox{hep-th/9905007}  }}
\title{Dilaton driven Hawking radiation in AdS$_2$ black hole}
\author{Won Tae Kim\footnote{electronic address:wtkim@ccs.sogang.ac.kr}
        and John J. Oh\footnote{electronic address:john5@cosmos.sogang.ac.kr}}
\address{Department of Physics and Basic Science Research Institute,\\
         Sogang University, C.P.O. Box 1142, Seoul 100-611, Korea}
\date{\today}
\maketitle
\bigskip 
\begin{abstract}
A recent study shows that Hawking radiation of a massless
scalar field does not appear on the two-dimensional AdS$_2$
black hole background. We shall study this issue by 
calculating absorption and reflection coefficients under
dilaton coupling with the matter field. 
If the scalar field 
does not couple to the dilaton, then it is fully absorbed
into the black hole without any outgoing mode. 
On the other hand, once it couples to
the dilaton field, the outgoing mode of
the massless scalar field exists, 
and the nontrivial Hawking radiation appears.
Finally, we comment on this dilaton dependence of Hawking
radiation in connection with 
a three-dimensional black hole.
\end{abstract}
\bigskip
\newpage
Recently, 
anti-de Sitter (AdS) spacetimes have been studied in connection with various aspects 
of AdS black hole physics, for example, in 
the calculation of black hole entropies 
\cite{ms1} in terms of lower-dimensional AdS gravity theories,
the three-dimensional Ba$\tilde{{\rm n}}$ados-Teitelboim-Zanelli(BTZ)
black hole \cite{btz} and AdS$_2$ black holes \cite{str1}.  

On the other hand, 
greybody factor \cite{dm,ms2} 
of the massless scalar field on this BTZ background
has been studied by Birmingham, Sachs, and Sen(BSS) \cite{bss} 
and they obtained the absorption cross section related to
the Hawking radiation \cite{haw} by carefully considering the
matching procedure between the bulk and the AdS 
boundary. In two-dimensional case, it has been shown
that the absorption 
coefficient is one and there are no reflection modes \cite{kop}, 
which means that there does not exist any massless radiation
on this two-dimensional AdS black hole background.
This fact seems to be consistent with the 
argument in Ref. \cite{kim} since expectation value of
the energy-momentum
tensor around the two-dimensional AdS black hole
is zero.  

In this paper, we would like to study 
this issue whether the Hawking radiation appears on the
two-dimensional AdS black hole background or not
in terms of a scattering analysis of the massless scalar
field as a test field. 
We shall assume that
the classical metric-dilaton background
and then the scalar wave equation will be solved
on this background for two cases.
For the first case, the free scalar field
equation is solved, and by carefully matching
this solution with the boundary solution we
obtain the expected null radiation.
Once the boundary condition that the outgoing
modes are absent is imposed at the horizon, 
there are
no more outgoing modes at the bulk and the boundary.   
On the other hand, for the second case of the
minimal coupling with the dilaton background, which is just a Jackiw-Teitelboim (JT) model \cite{jt},
the scalar field equation is exactly solved in the
bulk and the boundary.
By imposing appropriate boundary conditions, we obtain
at last the nontrivial Hawking radiation.
In this latter case, the vanishing limit
of the dilaton field does not exist in 
this calculation, and the two
cases are distinct in the scattering analysis.

Let us now assume the following metric-dilaton background as
\begin{eqnarray}
 ds^2 &=& - \left( -M + \frac{r^2}{\ell^2}  \right)
  dt^2 + \left( -M + \frac{r^2}{\ell^2}  \right)^{-1}
  dr^2, \label{metricsolution} \\
\psi& =& \gamma \ln \frac{r}{\ell} +\psi_0 \label{dilatonsolution},
\end{eqnarray}
where the metric (\ref{metricsolution}) 
describes AdS black hole with the horizon $r_{H}=\sqrt{M}\ell$
which is asymptotically AdS$_2$ spacetime,
and its curvature is a constant, $ R= - \frac{2}{\ell^2}$. 
One can assert that the Hawking temperature as 
\begin{equation}
  \label{ht}
  T_{H} = \frac{r_{H}}{2\pi \ell^2},
\end{equation}
which may be given by the conventional procedure to avoid 
the conical singularity of the metric in Euclidean formalism.
At first glance, this temperature, however, 
does not seem to depend on the dilaton
field. The purpose of the present paper in some sense is to
study the dilaton dependence of the black hole temperature
in terms of the scattering analysis.  

We are now going to study the scattering amplitudes
of a test field on the AdS black hole background in order to 
calculate the Hawking radiation and temperature from
the dynamical process.
Let us now consider the following scalar field $f$ obeying
\begin{equation}
\Box f + \nabla_{\mu} \psi \nabla^{\mu}f = 0 \label{matter}.
\end{equation} 
For $\gamma=0$, the matter field does not couple to the metric, 
whereas it couples explicitly with the dilaton field for $\gamma=1$. The Hawking radiation for the dilaton coupled scalar field
on the CGHS black hole background has been studied in Ref. \cite{no}. Here, we consider only two cases in order to compare
the behavior of scattering of the scalar field on the AdS black hole
background, and 
study the dilaton dependence of Hawking radiation.
In fact, for the case of $\gamma=0$ corresponding to
the constant dilaton background, 
the AdS$_2$ model \cite{kim} has been 
constructed by using the Callan-Giddings-Harvey-Strominger
(CGHS) model \cite{cghs} (especially, for the Russo-Susskind-Thorlacius model\cite{rst}, the AdS geometry was discussed in Ref.\cite{sol}). 
For the other case of $\gamma=1$, 
it has been well appreciated as a two-dimensional AdS
gravity \cite{cm,cad,lva}. 
Let us study Hawking radiation process on these backgrounds 
by calculating the absorption and reflection coefficients,
and obtain the desirable Hawking temperatures.  

After separation of variables with $f(t,r) = R_{(\gamma)}(r) e^{-i \omega t}$,
the spatial equation of motion yields
\begin{equation}
  \label{spatial}
  (r^2 - r_{H}^2 ) \partial_{r}^2 R_{(\gamma)}(r) + \frac{1}{r} 
\left[ 2 r^2 + \gamma (r^2 -
  r_{H}^2) \right]\partial_{r} R_{(\gamma)}(r) + \frac{\omega^2
  \ell^4}{(r^2 - r_{H}^2)} R_{(\gamma)}(r) = 0.
\end{equation}
Hereafter, let us treat two cases separately since they are
 drastically different. 

For the first
case of $\gamma=0$, we shall calculate the scattering
amplitude of the massless scalar field on the AdS black hole background, 
and then infer the Hawking radiation and temperature.  
By performing change of the variable as 
$z= \frac{r-r_{H}}{r+r_{H}}$ ($0 \le z \le 1$), 
the equation of motion Eq. (\ref{spatial}) 
can be simply written in the form of
\begin{equation}
  \label{a0}
  z(1-z) \partial_{z}^2 R_{(0)}(z) + (1-z) \partial_{z} R_{(0)}(z) +
 \frac{\omega^2 \ell^4}{4 r_{H}^2} (\frac{1}{z}-1) R_{(0)}(z) = 0.
\end{equation}
By defining $R_{(0)}(z) = z^{\alpha} g(z)$, 
the wave equation is given as
 \begin{equation}
  \label{singfeqna01}
   z(1-z) \partial_{z}^2 g(z) + (1-z)(2\alpha + 1) \partial_{z} g(z) +
   \left[ \frac{1}{z}\left( \alpha^2 + \frac{\omega^2 \ell^4}{4 r_{H}^2}
 \right) - \left(\alpha^2 + \frac{\omega^2 \ell^4}{4 r_{H}^2}
   \right) \right] g(z) = 0,
\end{equation}
and the nonsingular solution is finally obtained 
with $\alpha^2 =
-\frac{\omega^2 \ell^4}{4 r_{H}^2}$,
\begin{equation}
  \label{gensola0}
  R_{(0)}(r) = C_{{\rm in}} e^{-i\frac{\omega \ell^2}{2 r_{H}} \ln
  \left(\frac{r-r_{H}}{r+r_{H}}\right)} + C_{{\rm out}} 
e^{i\frac{\omega \ell^2}{2 r_{H}} \ln
  \left(\frac{r-r_{H}}{r+r_{H}}\right)}.
\end{equation}
Note that this is an exact solution defined over the bulk.  
Further, in the
far region, it is also simply written as 
\begin{equation}
  \label{farsola0}
  R_{(0)}^{{\rm far}}(r) = C_{\rm in} e^{i \frac{\omega \ell^2}{r}}
                         + C_{\rm out} e^{-i \frac{\omega \ell^2}{r}}.
\end{equation}
At this stage, we should carefully consider the boundary of this
AdS black hole because it has nontrivial background geometry in 
contrast to the asymptotically flat black hole. 
The background geometry of the usual black hole at
the asymptotically far region is happened to be 
that of the massless limit
of the black hole geometry.
So, in that case, the far region limit means 
the massless limit of the black hole geometry, 
however, in our model this is not the case.
Therefore, we should take the boundary metric by
defining $M=0$ in Eq. (\ref{metricsolution}). 
Then the equation of motion at the boundary 
is given by  
\begin{equation}
  \label{beqna0}
  r^2 \partial_{r}^2 R_{(0)}(r) + 2r \partial_{r}R_{(0)}(r) + \frac{\omega^2
  \ell^4}{r^2} R_{(0)}(r) = 0,
\end{equation}
and its solution is easily obtained as
\begin{equation}
  \label{boundarysola0}
 R_{(0)}^{\rm boundary}
(r) = A_{\rm in} e^{i \frac{\omega \ell^2}{r}} + A_{\rm out} e^{-i
  \frac{\omega \ell^2}{r}}.
\end{equation}
It is interesting to note that 
the asymptotic solution (\ref{farsola0}) is compatible with
the boundary solution (\ref{boundarysola0}) if we identify
$C_{\rm in}=A_{\rm in}$ and $C_{\rm out}=A_{\rm out}$. 

By using the expression of the flux expressed as
\begin{equation}
  \label{deflux}
  F_{(\gamma)} = \frac{2 \pi}{i} \left(\frac{r^2-
  r_{H}^2}{\ell^2}\right)\left[R^{*}_{(\gamma)}(r)\partial_{r}R_{(\gamma)}(r) -
 R(r)_{(\gamma)}
\partial_{r}R^{*}_{(\gamma)}(r)\right],
\end{equation}
we can straightforwardly define 
the absorption coefficient $(A)$ and reflection coefficient $(R)$ 
as 
\begin{equation}
  \label{absrefa0}
A=\left|\frac{F_{(\gamma)}^{\rm in}(r=r_H)}
           {F_{(\gamma)}^{\rm in}(r=\infty)}\right|,~~~~~~ 
R=\left|\frac{F_{(\gamma)}^{\rm out}(r=\infty)}
      {F_{(\gamma)}^{\rm in}
(r=\infty)}\right|,
\end{equation}
where $F_{(\gamma)}^{\rm in}(r=r_H; \infty)$ and
$F_{(\gamma)}^{\rm out}(r=r_H; \infty)$ are ingoing and outgoing fluxes at
the horizon and boundary, respectively.
At the horizon, we impose the boundary condition as
$C_{\rm out}=0$\cite{gl}, then the outgoing mode does not appear
at the bulk and the boundary. Therefore, the reflection coefficient
is zero, $R=0$. Even though one chooses the other boundary condition as $A_{\rm out}=0$ in Ref. \cite{gl}, in this case also the
reflection coefficient is zero as it should be. By the use of the following
relation which relates the reflection coefficient and Hawking
thermal radiation \cite{gl},
\begin{eqnarray}
  \label{hawka0}
  <0|N|0> = \frac{R}{1-R} = \frac{1}{e^{\frac{\omega}{T_H}}-1},
\end{eqnarray}
we can assume the vanishing Hawking temperature,
\begin{equation}
  \label{hawkrada0}
  T_H = 0
\end{equation}
for the finite mode of the scalar field, $\omega >0 $
since $R=0$.
This fact is compatible with the result of Ref. \cite{kim} 
where the Hawking radiation has been studied
in terms of the energy-momentum
tensor of massless scalar field.
Therefore, for $\gamma=0$ case, there is
no reflection mode in AdS$_2$ black hole background 
and there does not exist massless radiation in our
scattering analysis.

Let us now study the second case 
of $\gamma=1$ with the dilaton coupling of Eq. (\ref{matter}).
In this case, we perform a change of variable to solve
the field equation exactly as
$z=\frac{r^2 - r_{H}^2}{r^2}$ ($0 \le z \le 1$).
So the spatial(radial) equation of motion (\ref{spatial})
is given by 
\begin{equation}
  \label{speqna1}
  z(1-z)\partial_{z}^2 R_{(1)}(z) + (1-z)\partial_{z}R_{(1)}(z) +
  \frac{\omega^2 \ell^4}{4
  r_{H}^2 z} R_{(1)}(z) = 0,
\end{equation}
and it is rewritten as
\begin{equation}
    \label{preresula1}
     z(1-z)\partial_{z}^2 g(z) + 
    (1 + 2 \kappa )(1-z)\partial_{z} g(z)
     +  \left[ \frac{1}{z} \left(\kappa^2 
     +  \frac{\omega^2 \ell^4}{4 r_{H}^2}\right) - \kappa^2 \right]g(z) = 0,
\end{equation}
where $R_{(1)}(z) = z^{\kappa} g(z)$.
The equation of motion (\ref{preresula1}) becomes
\begin{equation}
    \label{speqna12}
    z(1-z)\partial_{z}^2 g(z)+(1+ 2 \kappa)(1-z)
\partial_{z}g(z) - \kappa^2 g(z) = 0
\end{equation}
after choosing the constant $\kappa$ as 
$\kappa^2 = - \frac{\omega^2
  \ell^4}{4 r_{H}^2}$,
and the standard solution is 
given by the linear combination of two hypergeometric
functions $F(a, b, c; z)$ and $z^{1-c}F(a+1-c, b+1-c, 2-c; z)$
where
\begin{equation}
  \label{coefa1}
  a = \kappa,~~ b = \kappa,~~ c = 1+ 2\kappa.
\end{equation}
Then, the bulk solution can be neatly written as
\begin{equation}
  \label{gensola1}
  R_{(1)}(r) =z^{-\kappa} C_{\rm in} F(-\kappa, -\kappa,
  1-2\kappa; z) + z^{\kappa} C_{\rm out} F(\kappa, \kappa,
  1+2\kappa; z).
\end{equation}
Note that it is symmetric under interchange of the sign of
$\kappa$, and we simply take the plus sign of $\kappa$.

In the near horizon limit ($z \rightarrow 0$), 
the solution is
reduced to
\begin{equation}
  \label{nearsola1}
  R_{(1)}^{\rm near}(r) = C_{\rm in} e^{-i \frac{\omega \ell^2}{2 r_{H}} 
\ln\left(\frac{r^2 -
  r_{H}^2}{r^2}\right)} + C_{\rm out} e^{i \frac{\omega \ell^2}{2 r_{H}} \ln\left(\frac{r^2 - r_{H}^2}{r^2}\right)}.
\end{equation}
On the other hand, 
the large-$r$ behavior of the bulk solution (\ref{gensola1}) follows
from the $z\rightarrow 1-z$ transformation law for the
hypergeometric functions \cite{as}:
\begin{eqnarray}
  \label{z1ztrans}
 F(a,b,a+b&+&m;z) =
  \frac{\Gamma(m)\Gamma(a+b+m)}{\Gamma(a+m)\Gamma(b+m)} \sum_{n=0}^{m-1}
  \frac{(a)_n (b)_n}{n! (1-m)_n} (1-z)^n \nonumber \\ 
&-& \frac{\Gamma(a+b+m)}{\Gamma(a)\Gamma(b)} (z-1)^{m}
  \sum_{n=0}^{\infty} \frac{(a+m)_n (b+m)_n}{n! (n+m)!}
  (1-z)^{n}\nonumber \\
&\times&  \left[\ln(1-z) - \psi(n+1) - \psi(n+m+1) + \psi(a+n+m) + \psi(b+n+m)\right].
\end{eqnarray}
Note that in our case $m$ is unity in Eq.(\ref{z1ztrans}).
Using this relation (\ref{z1ztrans}), we can obtain the far region solution
from Eq. (\ref{gensola1}), 
\begin{eqnarray}
  \label{farsola1}
  R_{(1)}^{\rm far}(r) &=& C_{\rm in} \frac{\Gamma(1-2
  \kappa)}{\Gamma(1 - \kappa)\Gamma(1-\kappa)} \left[ 1 + \kappa \left(\frac{r_{H}}{r}\right)^2 +
  \kappa^2 \left(\frac{r_{H}}{r}\right)^2
  \left(2 \ln\left(\frac{r_{H}}{r}\right) + \zeta_{-\kappa}^{(0)} \right)\right] \nonumber \\
&+& C_{\rm out} \frac{\Gamma(1+2
  \kappa)}{\Gamma(1+\kappa)\Gamma(1+\kappa)}\left[ 1 - \kappa \left(\frac{r_{H}}{r}\right)^2 + \kappa^2
  \left(\frac{r_{H}}{r}\right)^2 \left(2 \ln
  \left(\frac{r_{H}}{r}\right) + \zeta_{\kappa}^{(0)} \right)\right],
\end{eqnarray}
where $\zeta_{\pm \kappa}^{(0)} = 2 \psi(1\pm \kappa) - \psi(2)-\psi(1)$
and $\psi(z)$ is a digamma function. 

As was emphasized for the case of
$\gamma=0$, at the boundary $(z=1)$, we have to solve the wave
equation in order to match their coefficients with those of
Eq. (\ref{farsola1}). So the boundary wave equation
\begin{equation}
  \label{boundeqa1}
  r^2 \partial_{r}^2 R_{(1)}(r) + 3 r \partial_{r} R_{(1)}(r) +
  \frac{\omega^2
  \ell^4}{r^2} R_{(1)}(r) = 0,
\end{equation}
yields the following boundary solution
\begin{equation}
  \label{boundsola1}
  R_{(1)}^{\rm boundary}(r) =\frac
  {1}{r}\left[ \alpha  K_{1}\left(i\frac{\omega\ell^2}{r}\right) +
  \beta I_{1} \left(i\frac{\omega\ell^2}{r}\right)\right],
\end{equation}
where $K$ and $I$ are modified Bessel functions, and $\alpha$ and
$\beta$ are arbitrary constants. These functions are expanded as a
well-known form in Ref.\cite{as}, which is given by
\begin{eqnarray}
  \label{expbessel}
 I_{\nu} (z) &=& \left(\frac{1}{2}z\right)^{\nu} \sum_{k=0}^{\infty}
  \frac{\left(\frac{1}{2}z\right)^{2k}}{k! \Gamma(\nu+k+1)}, \nonumber
  \\
 K_{\nu} (z) &=& \frac{1}{2}\left(\frac{1}{2}z\right)^{-\nu}~
  \sum_{k=0}^{\nu-1} \frac{(\nu-k-1)!}{k!}
  \left(-\frac{z^2}{4}\right)^{k}\nonumber \\ &+& (-1)^{\nu+1} \ln
  \left(\frac{1}{2} z\right)I_{\nu}(z)\nonumber \\ &+&
 (-1)^{\nu}\frac{1}{2}
  \left(\frac{1}{2} z\right)^{\nu}\sum_{k=0}^{\infty}\left( \psi(k+1) +
  \psi(\nu+k+1)\right) \frac{\left(\frac{1}{2} z\right)^{2k}}{k!(\nu + k)!}.
\end{eqnarray}
Considering the most leading term which corresponds to $k=0$ in
Eq. (\ref{expbessel}) for the limit of
$r\rightarrow \infty$, we obtain the following boundary solution given
by
\begin{equation}
  \label{bsol2}
  R^{\rm boundary}_{(1)}(r) = a + \frac{b}{r^2},
\end{equation}
where $a$ and $b$ are arbitrary constants. We can rewrite this
solution in terms of the ingoing and outgoing modes by redefining the
constants as follows,
\begin{eqnarray}
  \label{redcoeff}
  a &=& A_{\rm in} + A_{\rm out}, \nonumber \\
  b &=& i\frac{{r_{H}}^2}{\pi}(A_{\rm in} - A_{\rm out}).
\end{eqnarray}
So we have the mode decomposed boundary solution by
\begin{equation}
  \label{bsoldecom}
  R_{(1)}^{{\rm boundary}}(r) = A_{\rm in} \left( 1 + i \frac{{r_{H}}^2}{\pi r^2}
  \right) + A_{\rm out} \left( 1- i \frac{{r_{H}}^2}{\pi r^2}\right).
\end{equation}
By matching the two solutions Eqs. (\ref{farsola1}) and
(\ref{bsoldecom}), the boundary coefficient $A_{\rm in}$ and $A_{\rm out}$ are determined as follows,
\begin{eqnarray}
  \label{matchco}
  A_{\rm in} &=& \frac{1}{2} C_{\rm out}
  \frac{\Gamma(1+2\kappa)}{\Gamma(1+\kappa) \Gamma(1+\kappa)} \left[ 1
  + i \pi \kappa \right]\nonumber\\ &+&
  \frac{1}{2} C_{\rm in} \frac{\Gamma(1-2\kappa)}{\Gamma(1-\kappa)
  \Gamma(1-\kappa)} \left[ 1 - i \pi \kappa \right], \nonumber \\
  A_{\rm out} &=& \frac{1}{2} C_{\rm out}
  \frac{\Gamma(1+2\kappa)}{\Gamma(1+\kappa) \Gamma(1+\kappa)} \left[ 1
  - i \pi \kappa \right]\nonumber\\ &+&
  \frac{1}{2} C_{\rm in} \frac{\Gamma(1-2\kappa)}{\Gamma(1-\kappa)
  \Gamma(1-\kappa)} \left[ 1 + i \pi \kappa \right].
\end{eqnarray}
Similarly to the previous case of $\gamma=0$, we choose the
boundary condition of $C_{\rm out}=0$ at the horizon\cite{gl},
and in the limit of small $\omega$, the absorption and reflection
coefficients are easily calculated as
\begin{eqnarray}
  \label{absrefa1}
  A=\left|\frac{F_{(1)}^{\rm in}(r=r_H)}
           {F_{(1)}^{\rm in}(r=\infty)}\right| &=& \frac{8 \omega
           \ell^2 r_{H}}{(2 r_{H} + \pi \omega \ell^2)^2},\nonumber \\
  R=\left|\frac{F_{(1)}^{\rm out}(r=\infty)}
      {F_{(1)}^{\rm in}
(r=\infty)}\right| &=& \left|\frac{A_{\rm out}}{A_{\rm in}}  \right|^2 
= \left(\frac{2 r_{H} - \pi \omega \ell^2}{2 r_{H} + \pi \omega \ell^2}
\right)
^2.
\end{eqnarray}

We now evaluate 
Hawking temperature by using the relation of reflection
coefficient and the number operator, 
\begin{eqnarray}
  \label{expeca1}
  < 0|N|0> & =& \frac{R}{1-R} \nonumber \\ 
               &=& \frac{1}{e^{\frac{\omega}{T_{H}}}-1}
\end{eqnarray}
where 
\begin{equation}
  \label{hawka11}
 T_H = \frac{\omega}{  \ln\left( 1 + \frac{8 \pi \omega \ell^2
 r_{H}}{(2 r_{H} - \pi \omega \ell^2)^2}\right)}.
\end{equation}
Note that if we take the $\omega \rightarrow 0$ limit, i.e., the
energy quanta of test field are so small, 
we obtain the desirable Hawking temperature
\begin{equation}
  \label{hawka12}
  T_{H} \approx \frac{r_{H}}{2 \pi \ell^2},
\end{equation}
which is interestingly coincident with the statistical Hawking
temperature Eq. (\ref{ht}). 
As for the second kind of boundary configuration, we 
can choose $A_{\rm out}=0$ as in \cite{gl}. 
So, after performing matching procedure,
we can calculate the ratio of coefficients between ingoing and
outgoing modes in the near horizon region, 
$  \left|\frac{C_{\rm in}}{C_{\rm out}}\right|^2 = 
 \left(\frac{2 r_{H} + \pi \omega \ell^2}{2 r_{H} - \pi \omega \ell^2}
\right)^2$.
We note that it is just the inverse of reflection
coefficient in the first kind of boundary condition
and the Hawking emission rate is given as
$ < 0|N|0 > = \left(\left|\frac{C_{\rm in}}
{C_{\rm out}}\right|^2 - 1\right)^{-1}$, which is in fact the same with
that of Eq. (\ref{expeca1}).

Note that 
we had an ambiguity in the boundary solution Eq. (\ref{bsoldecom})
in AdS$_2$ spacetime \cite{bss}, and the Hawking
temperature may not be uniquely determined.
This phenomenon has already appeared
in defining the surface gravity in the AdS black hole.
There is no preferred normalization of
timelike Killing vector in the asymptotic region
and the surface gravity depends upon this
normalization \cite{hw}.  So
we may not obtain the usual expression of the
Hawking temperature in our calculation too. 
However, in our boundary solution, we fixed
this ambiguity and the well-known expression
of the Hawking temperature Eq. (\ref{ht}) is obtained. 

We have shown that 
the nontrivial dilaton coupling for the second
case of $\gamma=1$ plays an important role in the Hawking radiation
process on the two-dimensional AdS black hole, which is 
contrasted with the first case of $\gamma=0$. 
This fact can be
traced back, and the origin may be
found from the three-dimensional BTZ black hole. Let us
consider the three-dimensional model as
\begin{equation}
  \label{3daction}
  S=\frac{1}{2\pi} \int d^3x \sqrt{-g} \left[ R + \frac{2}{\ell^2}
  \right] - \frac{1}{4\pi} \int d^3x \sqrt{-g} (\nabla f)^2
\end{equation}
which yields a well-known BTZ black hole solution
for $f=0$\cite{btz}. The absorption coefficient and the cross section
of the scalar field $f$ around 
the BTZ black hole background have been calculated
by BSS in Ref. \cite{bss}. Therefore 
if we take $S_1$ compactification along the angular axis as
$   {(ds)^2}_{(3)} = \left({g_{\alpha \beta}}^{(2)} +
   e^{2 \psi}A_{\alpha}A_{\beta}\right) dx^{\alpha}dx^{\beta} + 2
   e^{2 \psi} A_{\alpha}dx^{\alpha}dx_{3} 
   +e^{2 \psi} {dx_{3}}^2 $
where $\alpha,\beta = 0,1$, and $ x_{3}$ is 
angular coordinate in three dimensions, the
two-dimensional AdS black hole configuration  
of Eqs. (\ref{metricsolution}) and (\ref{dilatonsolution})
are obtained for the vanishing Kaluza-Klein charge \cite{str1}.
In this $s$-wave reduction, 
the equation of motion for the massless field
around this AdS$_2$ black hole becomes 
exactly Eq. (\ref{matter}) only for $\gamma=1$. 
Therefore, the
nonvanishing Hawking radiation in our model is due to the dilaton
related to the transverse radius of the 
three dimensions.  

On the other hand, 
the triviality of the Hawking radiation
for $\gamma=0$ is apparently 
due to the fact that the 
test field does not see the dilaton. In this case,
there may exist another explanation for this triviality
of the Hawking radiation. 
The constant $M$ in Eq. (\ref{metricsolution}) in 
the constant dilaton background can not be interpreted as a 
black hole mass as far as we take the background
metric as $ds^2 = - \frac{r^2}{\ell^2} dt^2 + 
\frac{\ell^2}{r^2} dr^2$ since the 
ADM mass is proportional to the derivative of 
the dilaton field.  To make this explicit, the quasilocal mass for the
asymptotically nonflat solution is given by
$Q_{\xi}=- \sqrt{-M +\frac{r^2}{\ell^2}}
\epsilon(r)$ where $Q_{\xi}$ is a conserved charge along the
timelike Killing vector $\xi$ and $\epsilon(r)$ is a local
energy density defined 
by $\epsilon(r)=-2\sqrt{-M +\frac{r^2}{l^2}}\partial_r e^{\psi}
-\epsilon_0(r)$ \cite{cmann}. 
The background energy is chosen
as $\epsilon_0 (r)=-\frac{2r}{\ell^2}$ to be satisfied with 
the condition of $\epsilon(r)=0$ for $M=0$.  
Then the ADM mass for the large$-r$ is given by
exactly $Q_{\xi}(r \rightarrow \infty)=\frac{M}{\ell}$ for
$\gamma=1$, while it becomes zero for 
the constant dilaton background case of $\gamma=0$.  
Therefore, in two dimensions, the nontrivial AdS black hole
solution seems to be accompanied with the nontrivial dilaton 
solution. 
This implies that for the case of $\gamma=0$,
the parameter $M $ in the metric Eq. (\ref{metricsolution})
may be removed by using some coordinate transformations,
which is in fact given by
$y^\pm =\frac{2l_{\rm eff}}{\sqrt{M}} {\rm tanh}
\frac{\sqrt{M}\sigma^\pm}{2l_{\rm eff}}$ where $y^\pm$ 
describes our vacuum geometry ($M=0$) of
$ds^2=-\frac{4l^2_{\rm eff}}{(y^+ - y^-)^2} dy^+ dy^-$ in
the conformal coordinate while $\sigma^\pm=t \pm r^*$ does 
the metric Eq. (\ref{metricsolution}).
The tortoise coordinate $r^*$ is defined by
$r^*= \int \frac{dr^*}{\sqrt{-M +\frac{r^2}{\ell^2}}}
=\frac{\ell^2}{2r_H} \ln \left( \frac{r-r_H}{r+r_H} \right)$. 
Therefore, one can think that the two-dimensional 
AdS black hole
and vacuum in some sense belong to the equivalent class
if the dilaton is not involved where the global difference between them
is rigorously studied very recently in Ref. \cite{ss}. 
Note that
for the dilaton coupled case of $\gamma=1$, the above
coordinate transformation is also possible, however, the
dilaton field in the new coordinate contains the information of
the AdS black hole.
Therefore the parameter $M$ turns out to be a coordinate
artifact for $\gamma=0$, and it seems to be meaningless 
to interpret it as a black hole mass in our calculation.

\vspace{20mm}

{\bf Acknowledgments}\\
This work was supported by Korea Research Foundation, 
No. BSRI-1998-015-D00074. 
We would like to thank M. Cadoni and D. Vassilevich for
many helpful comments.  

\end{document}